# Anomaly Detection Utilizing a Riemann Metric for Robust Myoelectric Pattern Recognition

ZongYe Hu, Ge Gao, Xiang Chen, *Member, IEEE*, and Xu Zhang, *Member, IEEE*

***Abstract*—Traditional myoelectric pattern recognition (MPR) systems excel within controlled laboratory environments but they are interfered when confronted with anomaly or novel motions not encountered during the training phase. Utilizing metric ways to distinguish the target and novel motions based on extractors compared to training set is a prevalent idea to alleviate such interference. An innovative method for anomaly motion detection was proposed based on simplified log-Euclidean distance (SLED) of symmetric positive definite manifolds. The SLED enhances the discrimination between target and novel motions. Moreover, it generates a more flexible shaping of motion boundaries to segregate target and novel motions, therefore effectively detecting the novel ones. The proposed method was evaluated using surface-electromyographic (sEMG) armband data recorded while performing 6 target and 8 novel hand motions. Based on linear discriminate analysis (LDA) and convolution prototype network (CPN) feature extractors, the proposed method achieved accuracies of 89.7% and 93.9% in novel motion detection respectively, while maintaining a target motion classification accuracy of 90%, outperforming the existing ones with statistical significance (p<0.05). This study provided a valuable solution for improving the robustness of MPR systems against anomaly motion interference.**

## I. INTRODUCTION

Surface electromyography (sEMG) signals, collected non-invasively from the skin surface, contain rich information for motor control [1]. It possesses the benefits of being non-invasive and portable [2]. Based on sEMG signals, the technique of myoelectric pattern recognition (MPR) decodes the gestural intentions by establishing a one-to-one mapping between the control commands and gesture patterns. Thus, MPR technology makes it available to control other devices dexterously, such as powered prostheses [1, 3, 4], exoskeletons [5-7], and mobile devices [8, 9].

The existing MPR systems achieved superior classification accuracy under ideal laboratory conditions through the combination of time-domain features and machine learning classifier [10-12]. However, their robustness still confronts challenges from various interferences that occur during daily usage, such as electrode shift [13-15], contraction strength changes [16, 17], variations in limb positions and orientation [10, 12], and novel motion interference [5, 18-26]. Among them, novel motion interference is a critical issue. The traditional MPR systems are prone to misclassify the novel motions as the target motions constrained within the closing-set during the training phase compulsorily [20, 21] and lead to mis-operation. Furthermore, the restriction in permissible motions is user-inhibitive, especially considering users' inadvertent engagement of novel motions in real scenarios. Consequently, novel motions are likely to degrade the performance of MPR systems and elicit users' negative feedback [22].

Many efforts have been conducted to alleviate the interference of novel motions [5, 18-26]. The prevalent idea is to establish a characterization threshold that delineates the distinction between the target and novel motions leveraging extracted features. This idea enables filtering out the motions beyond the threshold. Several characterizations are selected, such as probability [18, 23, 27], reconstruction error [19], and distance [20-22, 24-26].

The probability-based methods detect novel motions by examining the evidence of consistently low probability across all patterns. They have exhibited high accuracy in recognizing target motions using discriminative models such as support vector machines (SVM) [27], convolutional neural networks (CNN) [23], and linear discriminate analysis (LDA) [18]. However, its performance is interior in the detection of novel motions. The discriminative models simply allocate the feature space to the known/target classes in the closing training set, thereby resulting in the novel motions being recognized as the target ones under the criterion [20].

The probability-based methods reveal the exited contradiction between target motion recognition and novel one detection. To decouple the correlation, reconstruction methods are introduced. Reconstruction methods utilize autoencoders (AEs) [28, 29] to compress the feature of input motion and reconstruct it into the original form, thereby generating reconstruction error compared to the training set. The input motion is recognized as the novel one if the construction error exceeds the threshold. Reconstruction methods exhibit proficiency in both accurately detecting the novel motions and recognizing the target ones, attributed to the utilization of two

Manuscript Submitted 11 June 2024; revised xx xxxx 2024; accepted xx xxxx 2024. Date of publication xx xxxx 2024; date of the current version xx xxxx 2024. This work was supported in part the National Natural Science Foundation of China under Grant 62271464 and in part by Anhui Provincial Key Research & Development Program under Grant 2022k07020002. *(Corresponding author: Xu Zhang.)*

All authors are with Department of Electronic Science and Technology, University of Science and Technology of China, Hefei, Anhui 230027, P. R. China (e-mail: xuzhang90@ustc.edu.cn)

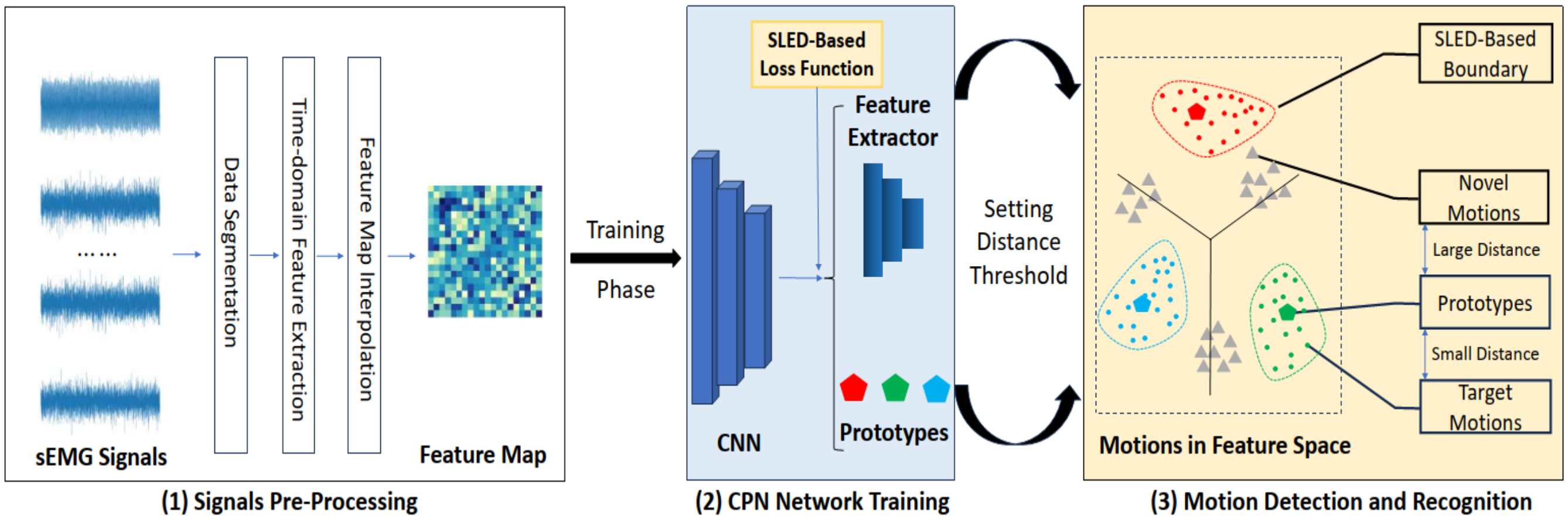

Fig. 1 Flowchart of the proposed method.

distinct networks tailored for varied purposes. However, the segmentation of networks elevates the burden of time complexity in the training phase and practical situation [19]

Distance-based methods integrated with generative models have been proposed to reinforce the performance in detecting novel motions [15, 20-22, 24-26]. These methods establish boundaries to enclose the different patterns of motions within specified regions in the feature space, facilitating novel motion detection through the assessment of whether an input motion is out of all the regions. Many previous studies reported classifiers integrated with Mahanobis Distance (MD) [21, 25, 26], support vector data description (SVDD) [15], and multiple one-versus-one classifiers [22, 24]. Distance-based methods enforce a more stringent criterion over probability-based methods, achieving superior performance in novel motion detection but inferior accuracy in recognizing the target ones. The distance-based and the derivative probability-based methods are widely favored considering the time burden and the requisite sample size under practical scenarios. The utilization is particularly prevalent following the optimization of training configurations based on metric learning facilitated the comparable performance to reconstruction-based methods [20].

It is crucial to make the MPR systems detect novel motions to alleviate their interference. However, while the above-mentioned methods achieve commendable efficiency in detecting novel motions, their enhancement primarily stems from optimization efforts in information mining strategies. Conversely, it is posited that effective metrics obtain the ability to reinforce the efficiency in data mining for their superior capability to describe the sample distribution [30]. Such an idea has exhibited noteworthy efficacy in augmenting the accuracy and robustness of many tasks such as skeleton-based hand gesture classification [31], image set classification [32], and face recognition [33], which share analogous characteristics with MPR tasks. With the above considerations, we propose an innovative method to yield the performance in novel motion detection based on a simplified log-Euclidean distance of symmetric positive definite manifolds (SLED), which is a kind of generalization of the CNN paradigm in distance metric to the dimension of Riemannian manifolds. The prevalent method based on metric learning is driven by the metric paradigm to evaluate the difference between different patterns. As a type of distance metric paradigm, the SLED is expected to improve differentiation between target and novel motions when integrated into metric learning-based networks during the training phase. Moreover, it could construct a more flexible boundary between the target and the novel motions by gaining more information from the distribution of the extracted features. Our study provides an efficient solution for detecting novel motions without a high computational burden in operation, enhancing the robustness of MPR system and its practical applicability.

## II. Methods

The flowchart of the proposed method is shown in Fig. 1. The time-domain features are first processed into feature maps. Then the feature maps together with the motion labels are used to train models for feature extraction, making better discrimination between different motion patterns through SLED-based loss function. In the testing phase, the SLED shapes a flexible boundary around the center of the target motions, engendering in a more effective detection of novel motions.

### A. Subjects

Eight able-bodied subjects participated in the study (aged from 21-26, right-handed). The Ethics Review Board of the University of the Science and Technology of China (Hefei, Anhui, China) approved the experimental protocol, under application No. 2022-N(H)-163, on February 2022. Each subject was provided with an explanation of the experiment's content and objectives, and written consent was obtained from each participant prior to their involvement.

### B. Experimental Procedure

During the experiment, the subjects were seated in chairs with their forelimbs resting comfortably on a table. The skin areas pertinent to the experiment were cleansed with rubbing alcohol.

Fourteen motions, including six target motions and eight novel motions, were investigated in the study. The known motions were radial deviation (T1), ulnar deviation (T2), wrist pronation (T3)/flexion (T4), and hand open (T5)/close (T6). These motions were adopted for their established significance in human-machine interaction. The novel motions comprised six static gestures: index extension (N1), pinch (N2), shoot (N3), OK (N4), and four daily tasks: handwriting (N5), mouse manipulating (N6), keyboard typing (N7), and scratch (N8). These motions were chosen due to their frequent occurrence in daily activities. Each motion was asked to be performed 15 times lasting for 2 seconds every time.

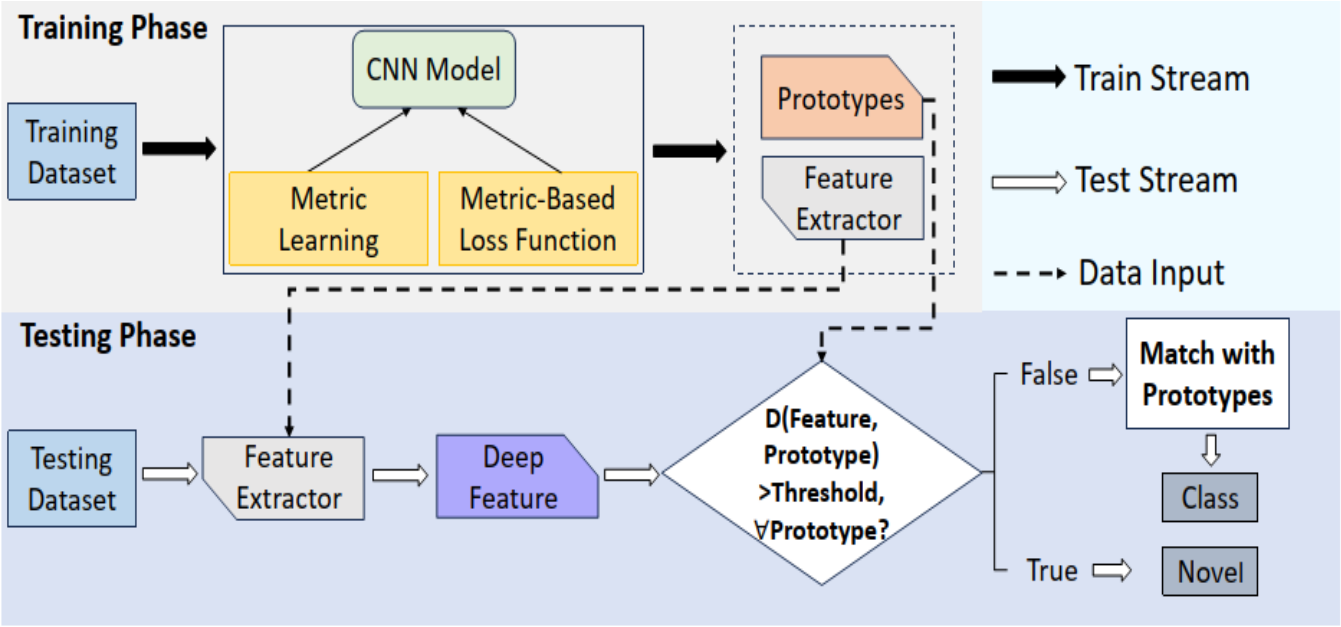

Fig. 2 Training and testing phases of the CPN.

### C. sEMG Pre-Processing

In this study, an 8-channel annular sEMG armband (gForce, OYMotion, China) is adopted for the acquisition of sEMG signals. The recorded signals are segmented into a succession of windows for analysis. Each window is 240ms in length and 80ms increments between consecutive windows.

Ten commonly used features are extracted for each analysis window, comprising four time-domain features [34] and six time-varying power spectrum descriptors [35]. These features possess the advantages of high efficiency for recognition and minimal computational burden [36]. Subsequently, each original feature map structured as $8 \times 10$ was expanded into $80 \times 80$ with the nearest interpolation to extend the receptive field of each layer.

### D. CPN Network

Convolution Prototype Network (CPN) is a modification of the network structure based on metric learning [33], which shows prominent efficiency in detecting novel motions by decreasing the variance within each pattern and enlarging the distance between different patterns [19, 20]. Fig. 2 demonstrates the training and testing phase of CPN. In the training phase, CPN learns a feature extractor $f(\boldsymbol{x},\theta)$ and prototype set $M=\{\boldsymbol{m_i}\}$, where $\boldsymbol{x}$ denotes the input feature maps, $\theta$ represents the parameters of the network, and $i \in \{1,2,\cdots k\}$ is the pattern of the target motions. In the testing phase, the input $\boldsymbol{x}$ belongs to the pattern $i$ only if the deep feature is close to the corresponding prototype $\boldsymbol{m_i}$ under defined metric $D(\cdot)$. Thus, the probability of a sample belonging to the prototype $\boldsymbol{m_i}$ can be defined as:

$$p(\boldsymbol{x} \in \boldsymbol{m}_i|\boldsymbol{x}) = \frac{e^{-D(f(\boldsymbol{x},\theta),\boldsymbol{m}_i)}}{\sum_{j=1}^{k} e^{-D(f(\boldsymbol{x},\theta),\boldsymbol{m}_j)}} \tag{1}$$

Based on the definition of $p(\boldsymbol{x} \in \boldsymbol{m}_i|\boldsymbol{x})$ and the idea of metric learning, the algorithm of CPN is formulated. It diverges from the traditional CNN models in the construction of loss function, which is defined as:

$$\begin{aligned}\ell_{total}\big((\boldsymbol{x},y)\colon\theta,M\big) &= \ell_{ce}\big((\boldsymbol{x},y)\colon\theta,M\big)\\ &+\lambda\ell_{pl}\big((\boldsymbol{x},y)\colon\theta,M\big)\end{aligned} \tag{2}$$

with each term defined as:

$$\ell_{ce}\big((\boldsymbol{x},y)\colon\theta,M\big) = -\ln\Big(p\big(\boldsymbol{x} \in \boldsymbol{m}_y|\boldsymbol{x}\big)\Big)$$

$$\ell_{pl}\big((\boldsymbol{x},y)\colon\theta,M\big) = D^2\big(f(\boldsymbol{x},\theta) - \boldsymbol{m}_y\big) \tag{3}$$

and $\lambda$ is a hyperparameter needed to be justified.

Fig 3 demonstrates the network structure of CPN adopted in this study. We train the network for 60 epochs with Adam algorithm utilized for optimization. The learning rate is initialized as $1.5e-4$ and dropped to one-tenth every 15 epochs.

### E. SLED

The metric defines the loss function utilized in the training phase as specified by equation (1-3), which operates on the output vector $f(\boldsymbol{x},\theta)$ and optimizes the paradigm of the network through backpropagation. Therefore, the selection of an appropriate metric is crucial for data mining. In this context, the Log-Euclidean Distance (LED) is employed.

Symmetric positive definite (SPD) manifolds $S(n)$ is composed of SPD matrices, which means they have positive eigenvalue only and can be defined as:

$$S(n) = \{S \in \mathbf{R}^{n\times n}, S = S^T : \boldsymbol{uSu^T} = 0, \forall \boldsymbol{u} \in \mathbf{R}^{1\times n}\ u \neq 0\} \tag{4}$$

In the SPD space, geometry metrics can be defined in various ways and the measurement can be deduced based on specific metrics. One metric is the Log-Euclidean Metric (LEM), which gives rise to the LED as the measurement of two different SPD matrices. The LED can be defined as:

$$LED^2(\boldsymbol{X},\boldsymbol{Y}) = Tr\left(\big(Ln(\boldsymbol{X}) - Ln(\boldsymbol{Y})\big)^2\right), \boldsymbol{X},\boldsymbol{Y} \in \boldsymbol{S}(n) \tag{5}$$

where the function $Tr(\cdot)$ denotes the trace of one matrix. The operation $Ln(\cdot)$ represents the Taylor expansion of matrix, with multiplication of matrices replacing the multiplication of scalar numbers, defined as:

$$\begin{aligned}Ln(\boldsymbol{X}) &= \boldsymbol{X} - \boldsymbol{I} - \frac{(\boldsymbol{X}-\boldsymbol{I})^2}{2} + \cdots + \frac{(-1)^{n-1}(\boldsymbol{X}-\boldsymbol{I})^n}{n} + \cdots\\ &= \boldsymbol{W^T}\begin{bmatrix}\ln(x_1) & \cdots & 0\\ \vdots & \ddots & \vdots\\ 0 & \cdots & \ln(x_n)\end{bmatrix}\boldsymbol{W}\end{aligned} \tag{6}$$

where matrix $\boldsymbol{W}$ is the eigenmatrix of $\boldsymbol{X}$ and $x_{1\cdots n}$ is the eigenvalue of $\boldsymbol{X}$.

To obtain symmetric positive definite matrices and reduce the burden of time complexity in the calculation of LED, Simplified Log-Euclidean Distance of symmetric positive definite manifolds (SLED) is proposed. The specific procedure will be delineated subsequently.

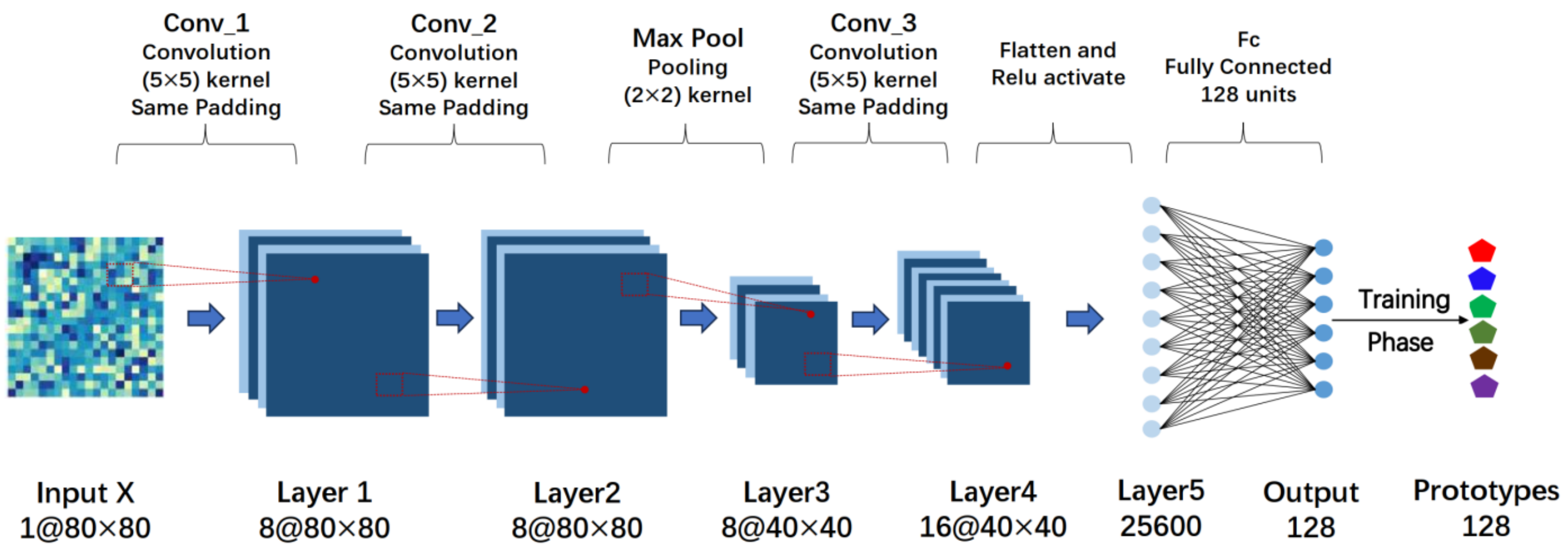

Fig. 3 Network structure of CPN adopted in this study. The network consists of 3 convolutional layers, 1 fully connected layer, and 1 max-pool layer.

Let $\boldsymbol{a}$, $\boldsymbol{b}$ denote two feature vectors extracted, with the shape of $1 \times n$ where $n$ represents the number of features. A symmetric positive matrix can be generated from $\boldsymbol{a}$ with the following process:

$$\boldsymbol{A} = \boldsymbol{a}^{\boldsymbol{T}}\boldsymbol{a} + \lambda \boldsymbol{I} \tag{7}$$

For $\boldsymbol{a}^{\boldsymbol{T}}\boldsymbol{a}$, we have:

$$\boldsymbol{a}^{\boldsymbol{T}}\boldsymbol{a}\boldsymbol{a}^{\boldsymbol{T}} = |a|^2 \boldsymbol{a}^{\boldsymbol{T}} \tag{8}$$

and

$$rank(\boldsymbol{a}^{\boldsymbol{T}}\boldsymbol{a}) \geq rank(\boldsymbol{a}^{\boldsymbol{T}}) + rank(\boldsymbol{a}) - n = 1$$
$$rank(\boldsymbol{a}^{\boldsymbol{T}}\boldsymbol{a}) \leq \min\left(rank(\boldsymbol{a}), rank(\boldsymbol{a}^{\boldsymbol{T}})\right) = 1 \tag{9}$$

which means $\boldsymbol{a}^{\boldsymbol{T}}\boldsymbol{a}$ have only one positive eigenvalue $|a|^2$ and others equal to 0. So $\boldsymbol{a}^{\boldsymbol{T}}\boldsymbol{a}$ can also be derived as:

$$\boldsymbol{a}^{\boldsymbol{T}}\boldsymbol{a} = \boldsymbol{U}^{\boldsymbol{T}} \begin{bmatrix} |a|^2 & \cdots & 0 \\ \vdots & \ddots & \vdots \\ 0 & \cdots & 0 \end{bmatrix} \boldsymbol{U} \tag{10}$$

where $\boldsymbol{U}$ is the eigenmatrix of $\boldsymbol{a}^{\boldsymbol{T}}\boldsymbol{a}$ and satisfies the conditions of:

$$\boldsymbol{U}^{\boldsymbol{T}}\boldsymbol{U} = \boldsymbol{I} \tag{11}$$

The first column of $\boldsymbol{U}^{\boldsymbol{T}}$ is equivalent to the normalization of $\boldsymbol{a}^{\boldsymbol{T}}$, denoted as the eigenvector corresponding to eigenvalue $|a|^2$, and is defined as $\boldsymbol{U}_{\boldsymbol{1}}^{\boldsymbol{T}}$. So $\boldsymbol{A}$ can be deduced as:

$$\boldsymbol{A} = \boldsymbol{a}^{\boldsymbol{T}}\boldsymbol{a} + \lambda \boldsymbol{I} = \boldsymbol{U}^{\boldsymbol{T}} \begin{bmatrix} |a|^2 & \cdots & 0 \\ \vdots & \ddots & \vdots \\ 0 & \cdots & 0 \end{bmatrix} \boldsymbol{U} + \lambda \boldsymbol{U}^{\boldsymbol{T}}\boldsymbol{U}$$
$$= \boldsymbol{U}^{\boldsymbol{T}} \begin{bmatrix} |a|^2 + \lambda & \cdots & 0 \\ \vdots & \ddots & \vdots \\ 0 & \cdots & \lambda \end{bmatrix} \boldsymbol{U} \tag{12}$$

And $\boldsymbol{b}$ obtains the same result as:

$$\boldsymbol{B} = \boldsymbol{V}^{\boldsymbol{T}} \begin{bmatrix} |b|^2 + \lambda & \cdots & 0 \\ \vdots & \ddots & \vdots \\ 0 & \cdots & \lambda \end{bmatrix} \boldsymbol{V} \tag{13}$$

$\boldsymbol{V}_{\boldsymbol{1}}^{\boldsymbol{T}}$ is the first column of eigenmatrix $\boldsymbol{V}^{\boldsymbol{T}}$.

Without loss of the generality, we can set $\lambda = 1$ to simplify the calculation. For the SLED between $\boldsymbol{A}$ and $\boldsymbol{B}$, we have:

$$SLED^2(\boldsymbol{A}, \boldsymbol{B}) = Tr\left(\left(Ln(\boldsymbol{A}) - Ln(\boldsymbol{B})\right)^2\right)$$
$$= Tr\left(\left(\begin{matrix} \boldsymbol{U}^{\boldsymbol{T}} \begin{bmatrix} \ln(|a|^2+1) & \cdots & 0 \\ \vdots & \ddots & \vdots \\ 0 & \cdots & 0 \end{bmatrix} \boldsymbol{U} \\ -\boldsymbol{V}^{\boldsymbol{T}} \begin{bmatrix} \ln(|b|^2+1) & \cdots & 0 \\ \vdots & \ddots & \vdots \\ 0 & \cdots & 0 \end{bmatrix} \boldsymbol{V} \end{matrix}\right)^2\right) \tag{14}$$

Let $a' = \ln(|a|^2 + 1), b' = \ln(|b|^2 + 1)$ for easy expression, we have:

$$SLED^2(\boldsymbol{A}, \boldsymbol{B}) = Tr\left(\begin{matrix} (a')^2 \boldsymbol{U}^{\boldsymbol{T}} \begin{pmatrix} 1 \\ \cdots \\ 0 \end{pmatrix} (1 \cdots 0) \boldsymbol{U} \\ +(b')^2 \boldsymbol{V}^{\boldsymbol{T}} \begin{pmatrix} 1 \\ \cdots \\ 0 \end{pmatrix} (1 \cdots 0) \boldsymbol{V} \\ -2a'b' \boldsymbol{U}^{\boldsymbol{T}} \begin{pmatrix} 1 \\ \cdots \\ 0 \end{pmatrix} (1 \cdots 0) \boldsymbol{U}\boldsymbol{V}^{\boldsymbol{T}} \begin{pmatrix} 1 \\ \cdots \\ 0 \end{pmatrix} (1 \cdots 0) \boldsymbol{V} \end{matrix}\right)$$
$$= Tr\left((a')^2 \boldsymbol{U}_{\boldsymbol{1}}^{\boldsymbol{T}}\boldsymbol{U}_{\boldsymbol{1}} + (b')^2 \boldsymbol{V}_{\boldsymbol{1}}^{\boldsymbol{T}}\boldsymbol{V}_{\boldsymbol{1}} - 2a'b'(\boldsymbol{U}_{\boldsymbol{1}}\boldsymbol{V}_{\boldsymbol{1}}^{\boldsymbol{T}})\boldsymbol{U}_{\boldsymbol{1}}^{\boldsymbol{T}}\boldsymbol{V}_{\boldsymbol{1}}\right) \tag{15}$$

For matrix $\boldsymbol{U}_{\boldsymbol{1}}^{\boldsymbol{T}}\boldsymbol{U}_{\boldsymbol{1}}$, we have:

$$Tr\left(\boldsymbol{U}_{\boldsymbol{1}}^{\boldsymbol{T}}\boldsymbol{U}_{\boldsymbol{1}}\right) = Tr\left(\begin{bmatrix} u_1u_1 & \cdots & u_1u_i & \cdots & u_1u_n \\ \vdots & \ddots & \ddots & \ddots & \vdots \\ \vdots & \ddots & u_iu_j & \ddots & \vdots \\ \vdots & \ddots & \ddots & \ddots & \vdots \\ u_nu_1 & \cdots & u_nu_i & \cdots & u_nu_n \end{bmatrix}\right)$$
$$= \boldsymbol{U}_{\boldsymbol{1}}\boldsymbol{U}_{\boldsymbol{1}}^{\boldsymbol{T}} \tag{16}$$

This property also applies to $\boldsymbol{V}_{\boldsymbol{1}}\boldsymbol{V}_{\boldsymbol{1}}^{\boldsymbol{T}}$ and $\boldsymbol{U}_{\boldsymbol{1}}\boldsymbol{V}_{\boldsymbol{1}}^{\boldsymbol{T}}$. Furthermore, since the operation $Tr(\cdot)$ obeys distributive law, $SLED^2(\boldsymbol{A}, \boldsymbol{B})$ can be simplified as:

$$SLED^2(\boldsymbol{A}, \boldsymbol{B}) = (a')^2 \boldsymbol{U}_{\boldsymbol{1}}\boldsymbol{U}_{\boldsymbol{1}}^{\boldsymbol{T}} + (b')^2 \boldsymbol{V}_{\boldsymbol{1}}\boldsymbol{V}_{\boldsymbol{1}}^{\boldsymbol{T}} - 2a'b'\left(\boldsymbol{U}_{\boldsymbol{1}}\boldsymbol{V}_{\boldsymbol{1}}^{\boldsymbol{T}}\right)^2 \tag{17}$$

And because $\boldsymbol{U}_{\boldsymbol{1}} = \frac{\boldsymbol{a}}{|\boldsymbol{a}|}$, $\boldsymbol{V}_{\boldsymbol{1}} = \frac{\boldsymbol{b}}{|\boldsymbol{b}|}$, we have:

$$SLED^2(\boldsymbol{A}, \boldsymbol{B}) = (a')^2 + (b')^2 - \frac{2a'b'}{|\boldsymbol{a}|^2|\boldsymbol{b}|^2}(\boldsymbol{ab})^2 \tag{18}$$

In this way, the time complexity of LED is reduced to $O(n)$ equal to Euclidean Distance (ED), where only dot between vectors and other basic operations are involved.

### F. Novel Motion Detection

In the testing phase, the predicted label $y_{predict}$ of the input map $\boldsymbol{x}$ can be determined based on the probability defined in equation (1), presented as:

$$y_{predict} = \arg\max_{i\in\{1\cdots k\}} p(\boldsymbol{x} \in \boldsymbol{m}_i|\boldsymbol{x}) \quad (19)$$

The condition is equivalent to:

$$y_{predict} = \arg\min_{i\in\{1\cdots k\}} D(\boldsymbol{x} \in \boldsymbol{m}_i|\boldsymbol{x}) \quad (20)$$

Furthermore, as the distance metric possesses the capacity to elucidate the difference between the corresponding prototype and samples, a threshold $T$ can be established to facilitate novel motion detection.

$$y_{novel} = \begin{cases} 0, D\left(f(\boldsymbol{x},\theta), \boldsymbol{m}_{\boldsymbol{predict}}\right) \le T \\ 1, D\left(f(\boldsymbol{x},\theta), \boldsymbol{m}_{\boldsymbol{predict}}\right) > T \end{cases} \quad (21)$$

Therefore, we can detect novel motions by predicting the label of one motion initially and then comparing the distance between the motion and the corresponding prototype with the threshold $T$.

We implemented all algorithms discussed using Python language and PyTorch architecture, and ran them on a laptop computer with an Intel i7 CPU and NVIDIA GeForce RTX 3060 laptop GPU.

### G. Performance Evaluation

Two evaluation ways were utilized to assess the performance of the proposed method designed for detecting novel motions and recognizing the target ones. They are the receiver operating characteristics (ROC) curve[37] and the classification accuracy.

As an evaluation for binary classifiers, the ROC reflects the performance of the proposed method in detecting the novel motions and recognizing the target ones respectively. The curve plots the true positive rates (TPR) and the false positive rates (FPR) across different thresholds. Among this, the TPR measures the ratio of target motions being recognized correctly, while the FPR measures the ratio of novel motions being identified as the target ones incorrectly. The area under the curve (AUC) is usually considered as an indicator evaluating the overall performance of one classifier, and a higher AUC indicates better performance for a binary classifier.

Following the establishment of the threshold, we can also calculate the accuracy of each class. The accuracy is the ratio of samples that are identified correctly for each pattern. The proposed method was evaluated in a user-specific manner and all the resultant indicators were averaged across 5-fold cross-validations.

To evaluate the performance of the proposed methods, several comparison methods are implemented.

1) **CPN integrated with ED (CPN-ED)**[20]. This algorithm has reported superior performance in novel motion detection. It utilizes ED in equations (1-3) and (19) to substitute SLED for target motion recognition and novel one detection.
2) **LDA integrated with Mahalanobis Distance (LDA-MD)**[25, 26]. This is a traditional algorithm for novel motion detection, served as the baseline method in this study. LDA is a traditional algorithm for MPR systems. It degrades the dimension of the input vectors to augment the distance between different patterns and decrease the variance within each class. And because of this, LDA serves the function aligned with CPN. Thus, the distance can also indicate the difference between samples and prototypes. Precisely, the prototype set $M = \{\boldsymbol{m_i}\}$ can be defined as:

$$\boldsymbol{m_i} = \overline{\sum_{j=1,y_j=i}^{k} \boldsymbol{x_j}} \quad (22)$$

And MD can be deduced as:

$$MD\left(\boldsymbol{x}, \boldsymbol{m}_{y_x}\right) = \left(\boldsymbol{x} - \boldsymbol{m}_{y_x}\right)\sigma_{y_x}^{-1}\left(\boldsymbol{x} - \boldsymbol{m}_{y_x}\right)^T \quad (23)$$

where the $\boldsymbol{x}$ denotes the input vector, $\boldsymbol{m}_{y_x}$ is the prototype in equation (22), and the matrix $\sigma_{y_x}$ is the covariance matrix of the training data in pattern $y_x$. The method detects novel motions with the criterion in equation (21).
3) **LDA integrated with SLED (LDA-SLED)**. To demonstrate that SLED also yields performance based on traditional feature extractors, LDA-SLED is proposed. The method utilizes SLED to replace MD in novel motion detection.
4) **LDA integrated with ED (LDA-ED)**. It serves the parallel algorithm with LDA-MD by replacing the MD with ED, both of which are straightforward metrics of distance.

### H. Statistics Analysis

Two One-way repeated-measure ANOVAs were utilized to compare the AUC and classification accuracy of different methods (LDA-MD, LDA-ED, LDA-SLED, CPN-ED, CPN-SLED), respectively. The significant level was set to 0.05. SPSS software was used for analysis. (ver26.0, SPSS Inc. Chicago, IL, USA)

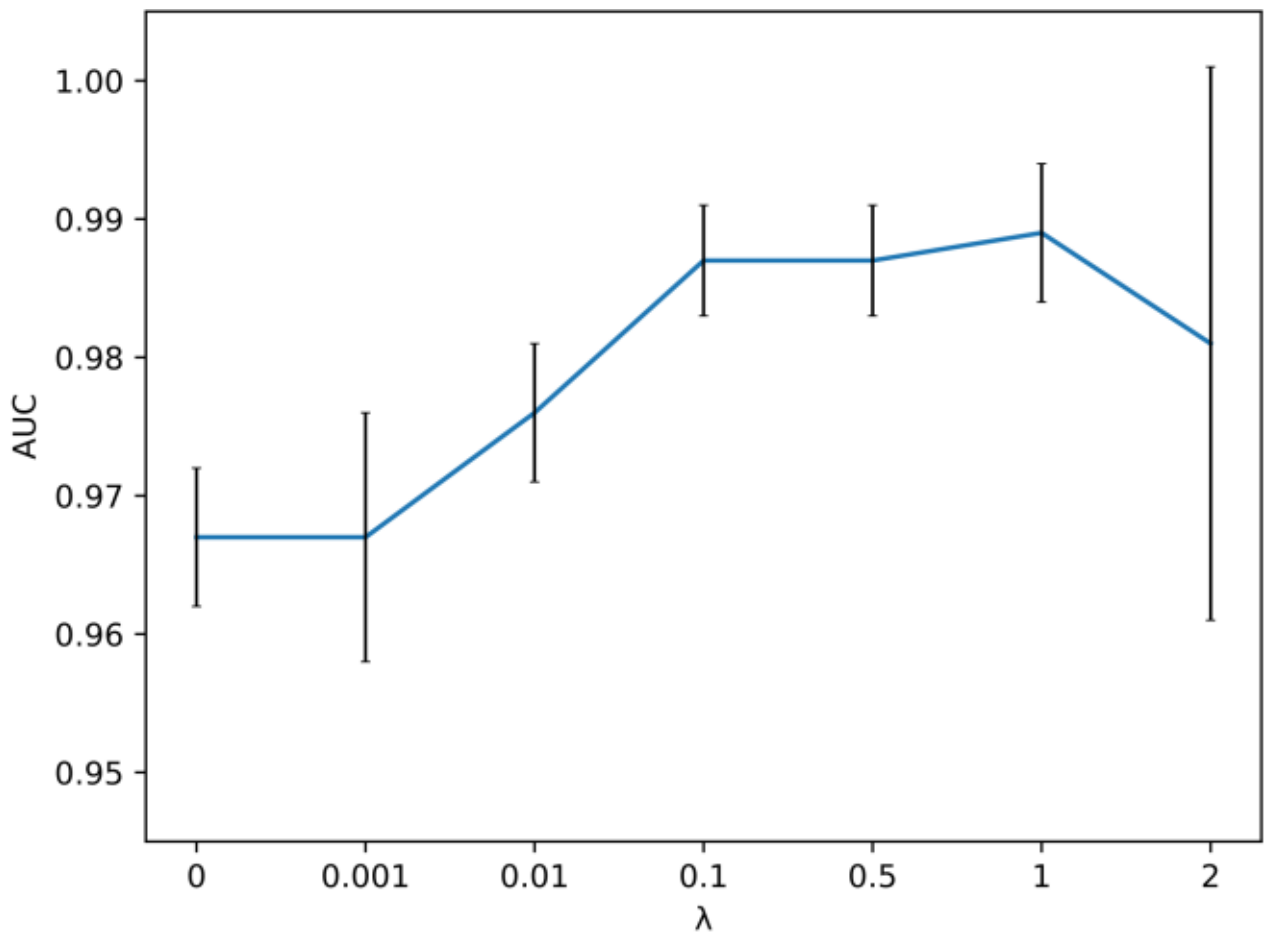

Fig. 4 AUC value achieved by the proposed method in one user with different λ. Error bar represents standard error generated by 5-fold cross-validations.

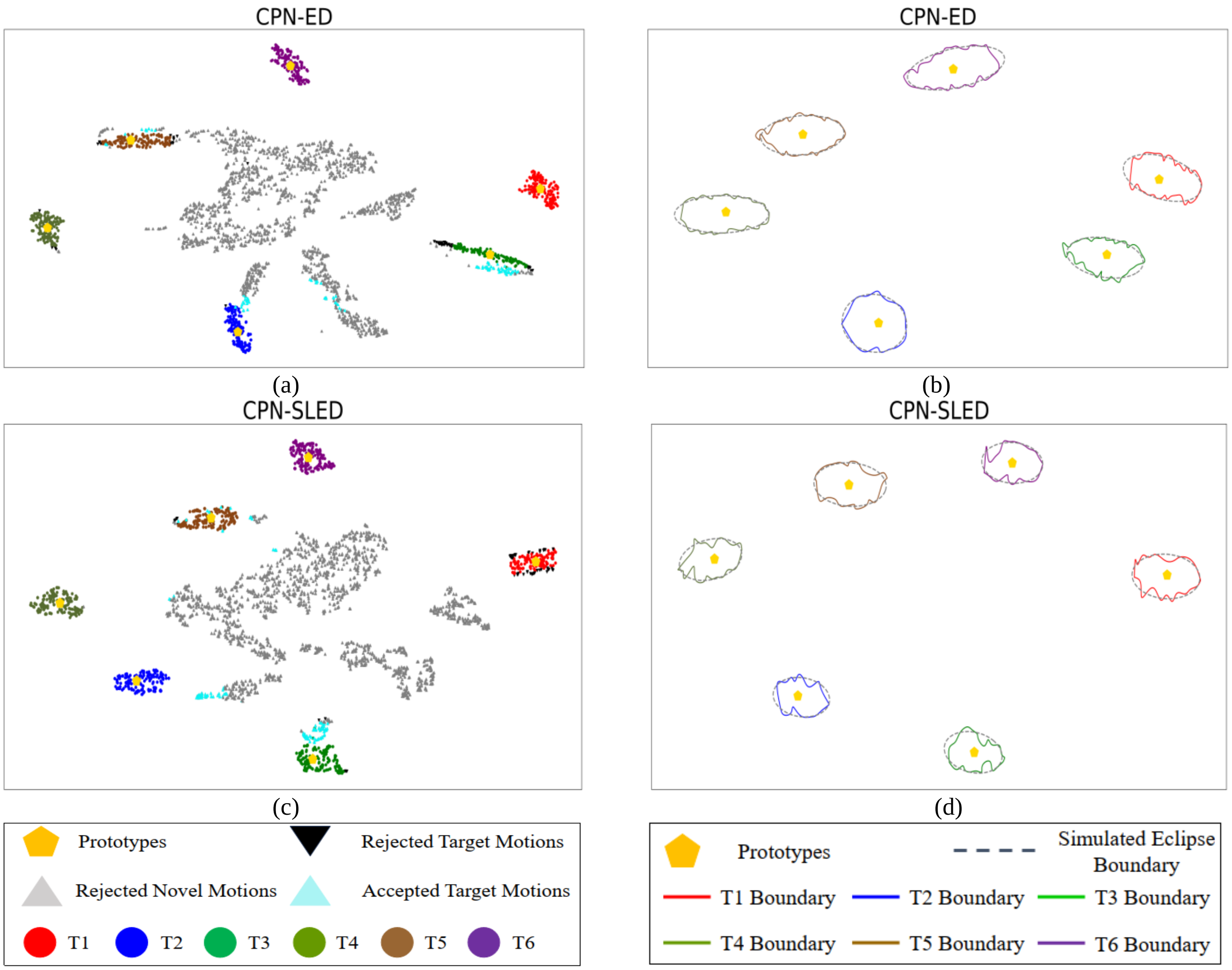

Fig. 5 t-SNE visualization of the distribution and the boundary of different motion patterns in feature space. (a), (c) refer to the distribution of different patterns. (b), (d) refer to the boundary around different prototypes. All the boundaries in (b), (d) are simulated through random points around the real boundary. (a), (c) and (b), (d) held same TPR respectively. All the data is collected from the user whose data is used in the training phase.

## III. Results

To examine the performance of the CPN-based methods under different λ and engender a more robust basis for comparison, Fig. 4 delineates the AUC plotted against λ for CPN-SLED utilizing training data from a singular user. The experimental protocol adheres to the 5-fold cross-validation strategy. Notably, the AUC rises while λ lies in the interval from 0 to 1.0 and attains its zenith when λ equals 1.0. This was always the case for data from other subjects. Thus, λ was held at 1.0 for subsequent analytical endeavors. Conversely, λ was held at 0.5 based on the same standard for the CPN-ED method.

Fig. 5 presents a graphical representation of disparate labels within the feature space utilizing t-distributed stochastic embedding (t-SNE) [38] technology. It is clear that CPN-SLED exhibits fewer misclassifications in novel motion detection compared to CPN-ED while upholding commensurate accuracy in recognizing target motions. Besides, a more noticeable disparity around the prototypes and flexible boundary can be noticed in Fig. 5.

To facilitate a meticulous comparison aimed at the user whose data was used for λ justification, Fig. 6 presents the confusion matrices of both the CPN-ED and CPN-SLED methods. The confusion matrices document the proportion of correctly classified motions for each category. Notably, the results indicate that CPN-SLED facilitates a higher detection rate of novel motions across all categories. This effect is particularly pronounced in the case of motions such as N1 and N3, wherein CPN-ED demonstrates comparatively inferior performance. Besides, the accuracy of different kinds of target motions in CPN-SLED is more even than in CPN-ED, which exhibits an especially low recognition rate in T3.

Fig. 7 illustrates the average ROC curves for the five methods considered. These graphical representations elucidate the comprehensive performance across various threshold values (T), with the desirable curve tending toward the upper-left corner. Among the different methods groupings, those incorporating SLED achieve 0.968 for CPN-SLED and 0.949 for LDA-SLED, exhibiting notably higher AUC values respectively ($p<0.05$). Additionally, a significant discrepancy in FPR, denoting the ratio of the novel motions recognized as the target ones, underscores the superior performance of classifiers employing SLED over their counterparts, particularly under strict settings that TPR is constrained at 0.9.

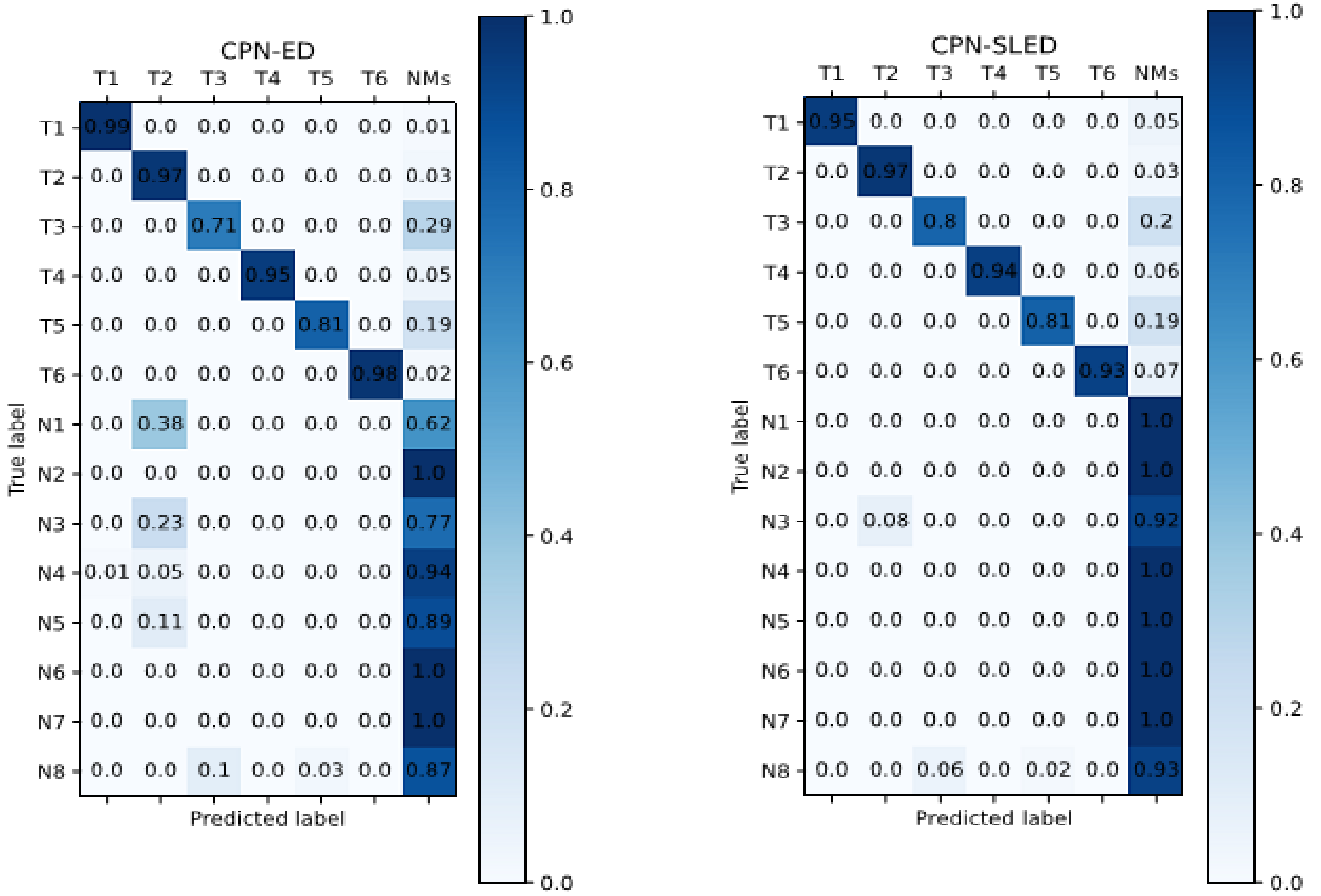

Fig. 6 Confusion matrices of CPN-ED and CPN-SLED of the user for hyperparameter λ justification.

For a more detailed comparison among the various methods, Fig. 8 analyzes the difference between the FPR. Noteworthy is the observation that the methods integrated with SLED yield accuracy in novel motion detection with statistical significance ($p<0.05$), achieving 93.9% and 89.7% under CPN and LDA feature extractors respectively. Particularly it could be noticed that the enhanced performance of extractors employing SLED in motions such as N1, N3, and N4, which are challenging to detect, is higher than the traditional metrics integrated with different classifiers respectively. In these instances, classifiers incorporating SLED exhibit a discernible superiority over comparison methods, with statistical significance ($p<0.05$).

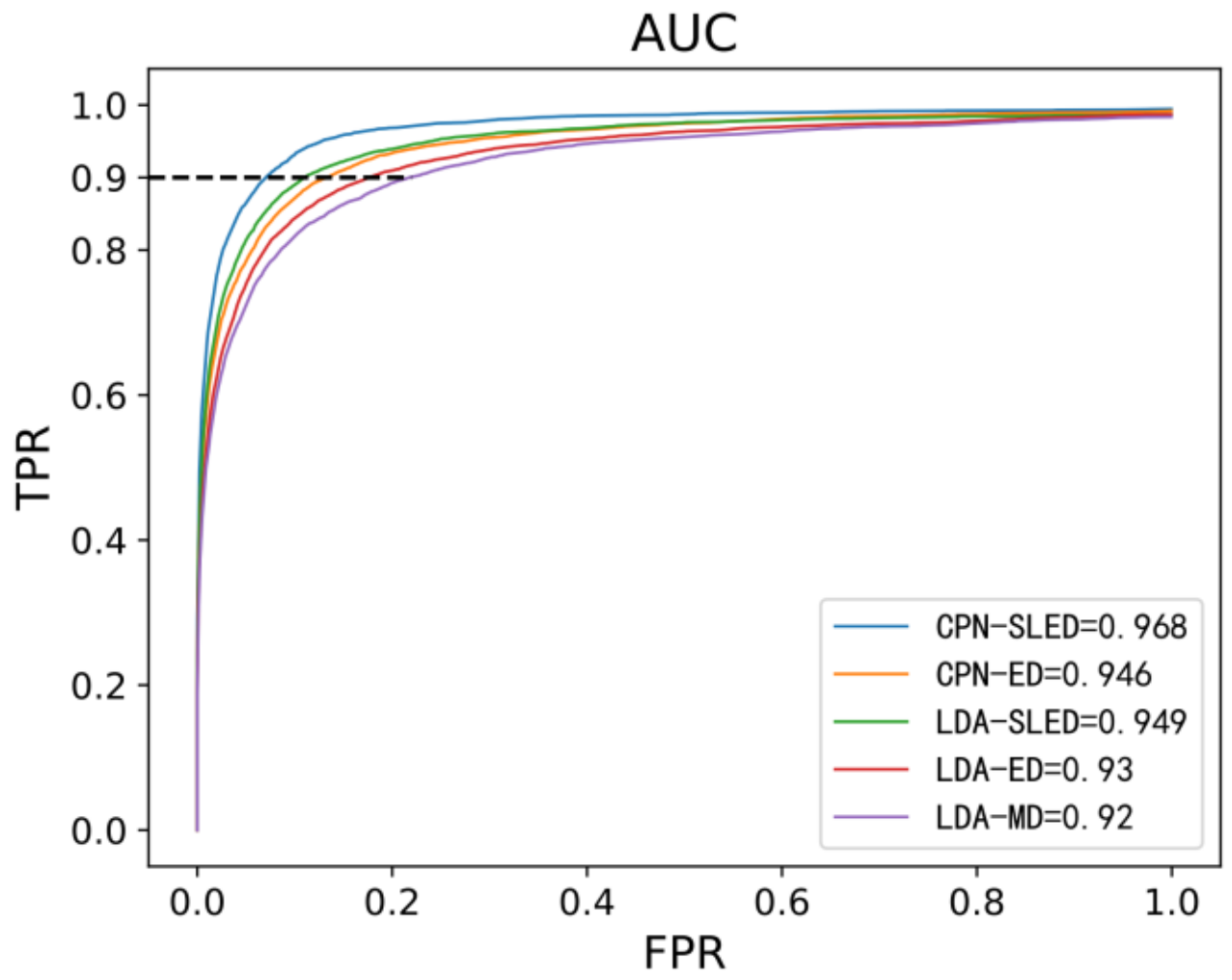

Fig. 7 ROC curve across five methods over all subjects

## V. Discussion

The capacity to effectively detect novel motions is paramount for enhancing the robustness of MPR systems. This paper presents an innovative method to effectively distinguish target and novel motions, without imposing an extra time complexity burden in the novel motion detection process.

SLED possesses the property that augments the isolation between target and novel motions, thereby facilitating superior novel motion detection. As depicted in Fig. 5, a noticeable disparity is observed in the prototype of target motions relative to the cluster of novel motions, particularly evident around T2. This discrepancy can be attributed to variations in the metric way utilized during the training phase. SLED comprehensively mines differences and similarities during the training phase, resulting in more notable isolation between target and novel motions. These findings substantiate the feasibility of the proposed method.

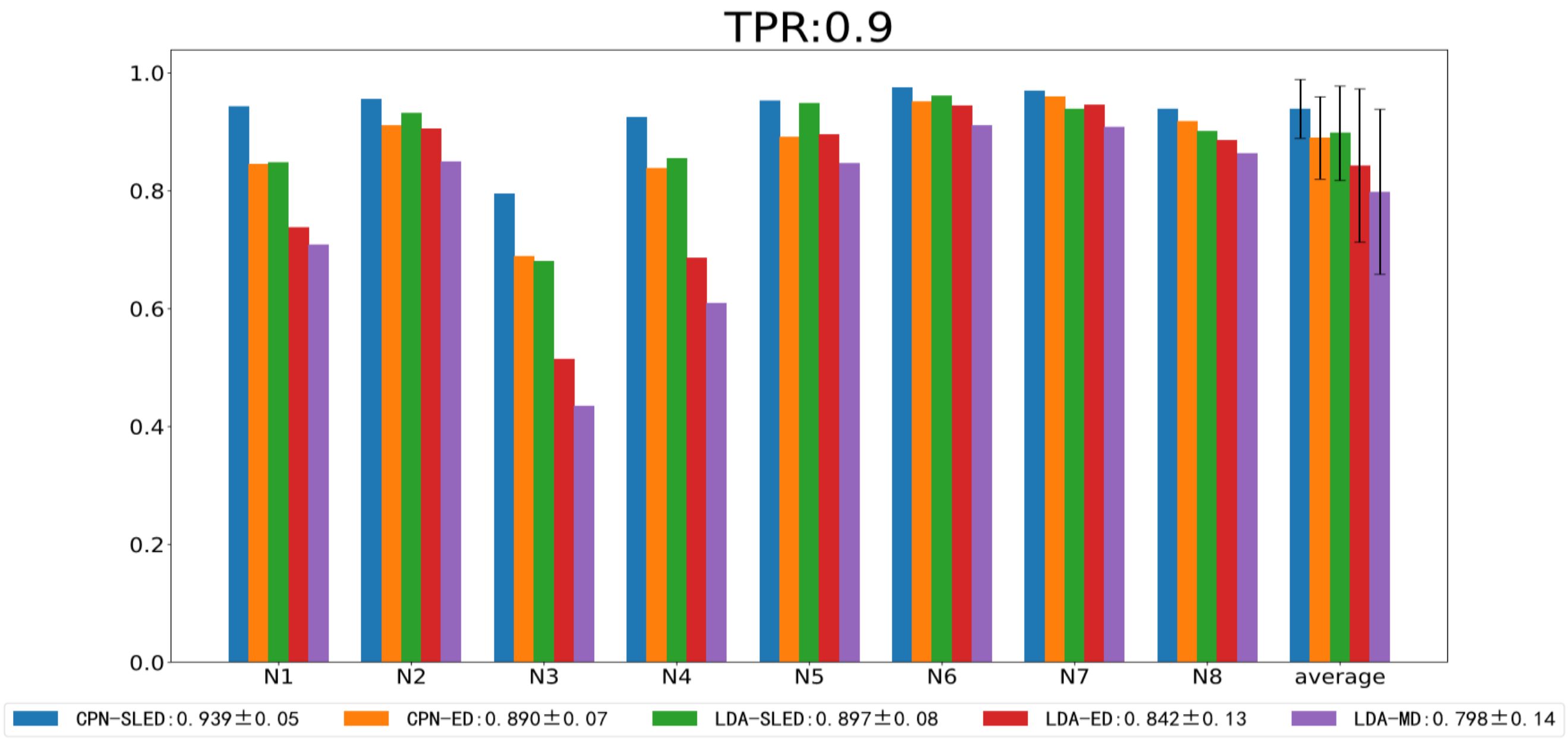

Fig. 8 Accuracy in novel motion detection across five methods over all subjects. The threshold is justified to make TPR equal to 0.9, consistent with the formal condition.

Utilizing SLED leads to the establishment of a more flexible boundary, resulting in a reduction in the area enclosed within the boundary and thereby enhancing the ability to detect novel motions. Notably, Fig. 5 illustrates that fewer novel motions located farther from the prototype center are misclassified as target motions with CPN-SLED, indicating a smaller enclosed boundary area. The difference in efficiency is attributed to the shape of the boundary generated by different metric ways. The boundary generated by ED assumes a high-dimensional spherical shape, indicating isotropy [40]. Consequently, the farthest samples among all the directions dictate the boundary radius, even if the radius exceeds the range required to encompass motions from other directions. Fig. 5(b) illustrates that the boundary generated by CPN-ED is more similar to the eclipse simulated, indicative of the projection of a high-dimensional sphere [40]. The result is consistent with the previous research [21]. Conversely, SLED incorporates more information from the distribution of deep features, resulting in a flexible boundary, which matches the previous research [39], and a smaller enclosed area. In this way, the SLED boundary is more flexible than the eclipse simulated, resulting in a relatively small enclosed area that detects more novel motions. For instance, boundary T1 in Fig. 5(d) exhibits a heart-like shape, diverging from the elliptical or spherical shape characteristic of high-dimensional ball projections. Visualization of motions in the feature space demonstrates the efficacy of the proposed method.

Moreover, SLED exhibits universality in achieving enhanced performance across various classifiers as depicted in Fig. 8. The process by which SLED assimilates information from the extracted feature distribution and delineates the boundary is independent of the involvement of SLED during the training phase and the classifier type. Besides, a minimal difference is observed between the AUC of LDA-ED and LDA-MD. It is attributable to the high-dimension elliptical boundary shaping by MD [21] similar to the high-dimensional ball configuration facilitated by ED in certain scenarios, resulting in comparable areas within the boundary. The university for different classifiers underscores SLED's adaptability, suggesting its potential applicability beyond the structure of MPR systems.

Significantly, methods with SLED consistently achieve superior accuracy in detecting any kind of novel motion under identical classifiers. As evidenced in Fig. 8, SLED consistently outperforms comparison methods across all novel motion types, with substantial disparities observed in certain motions (N1, N3, N4), while others are more marginal. The prevalent enhancement in efficiency attributable to SLED stems from the refinement of the metric way, independent of motion type. Consequently, the more effective delineation of differences between novel and target motions is a universal attribute across all novel motion types. Furthermore, distinctions among different types of novel motions primarily hinge on the similarity of samples to prototype centers. Samples proximate to prototypes are enclosed within the boundary for all methods, while those distant are more readily detected due to the shrinking area encapsulated by the boundary established through SLED. This universality in achieving heightened performance across all novel motion patterns, coupled with relatively low operational time complexity, underscores the potential of SLED to replace traditional metric ways, thereby bolstering execution robustness.

However, limitations exist in this study still. Only one user's data is used for justifying the hyperparameter $\lambda$ to modify the practical situation that models are justified on few data collected. Ongoing work is to enlarge the scale of the dataset and find a better ratio of data used for model justifying. Besides, the exited structure of networks is constructed and optimized based on Euclidean space, descending the performance of SLED in the loss function. Current work is to reconfigure different layers of the network and make the usage of SLED more effective. It is hoped that the time burden can be lowered and the efficiency of detecting novel motions. All these topics will be the directions of the future work.

## V. Conclusion

This paper presents an innovative metric scheme for elevating the robustness of MPR system against the interference of novel motions. The proposed method gained more information from the dataset and shaped a flexible boundary of the target motions in feature space. Thus, the method achieves higher accuracy of novel motion detection in CPN and LDA feature extractors respectively under the same setting of the accuracy in recognizing the target motions. The results indicate that the proposed method is promising in enhancing the robustness of MPR systems.